\documentclass[11pt]{article}
\usepackage{epsfig}

\begin{document}
\renewcommand{\thefootnote}{\fnsymbol{footnote}}

\begin{center} 
{\Large \bf Topological Structure of Dense Hadronic Matter\footnote{Talk 
given at the KIAS-APCTP Symposium in Astro-Hadron Physics "{\it Compact Stars: 
Quest for New States of Dense Matter\/}," November 10-14, 2003, Seoul, Korea} }
\\
\vspace{.30cm}
Byung-Yoon Park$^{a}$, Hee-Jung Lee$^{b}$, Vicente Vento$^{b}$, \\
Joon-Il Kim$^{c}$, Dong-Pil Min$^{c}$ and Mannque Rho$^{d,e}$

\vskip 0.20cm

{(a) \it Department of Physics,
Chungnam National University, Daejon 305-764, Korea}\\
({\small E-mail: bypark@cnu.ac.kr})

{(b) \it Departament de Fisica Te\`orica and Institut de
F\'{\i}sica Corpuscular}\\
{\it Universitat de Val\`encia and Consejo Superior
de Investigaciones Cient\'{\i}ficas}\\
{\it E-46100 Burjassot (Val\`encia), Spain} \\ 
({\small E-mail: hjlee@phya.snu.ac.kr, Vicente.Vento@uv.es})

{(c) \it Department of Physics,
Seoul National University, Seoul 151-742, Korea}\\
({\small E-mail: jikim@phya.snu.ac.kr,dpmin@phya.snu.ac.kr})

{(d) \it School of Physics, Korea Institute for Advanced Study,
Seoul 130-722, Korea}

{(e) \it Service de Physique Th\'eorique, CEA Saclay}\\
{\it 91191 Gif-sur-Yvette, France}\\
({\small E-mail: rho@spht.saclay.cea.fr})

\end{center}
\vskip 0.3cm

\centerline{\bf Abstract} 
We present a summary of work done on dense
hadronic matter, based on the Skyrme model, which provides a
unified approach to high density, valid in the large $N_c$ limit.
In our picture, dense hadronic matter is described by the {\em
classical} soliton configuration with minimum energy for the given
baryon number density. By incorporating the meson fluctuations on
such ground state we obtain an effective Lagrangian for meson
dynamics in a dense medium. Our starting point has been the Skyrme
model defined in terms of pions, thereafter we have extended and
improved the model by incorporating other degrees of freedom such
as dilaton, kaons and vector mesons. 

\renewcommand{\thefootnote}{\arabic{footnote}}
\setcounter{footnote}{0}

\section{Introduction}
At high temperature and/or density, hadrons are expected to
possess properties that are very different from those at normal
conditions. Understanding the properties of hadrons in such
extreme conditions is currently an important issue not only in
nuclear and particle physics but also in many other related fields
such as astrophysics. Data from the high energy heavy ion
colliders, astronomical observations on compact stars and some
theoretical developments have shown that the phase diagram of
hadronic matter is far richer and more interesting than initially
expected. Lattice QCD calculations have been carried out
successfully at high temperature, however similar calculations at
high density have not yet been possible\cite{lattice03}.
Theoretical developments have unveiled such interesting QCD phases
as color superconductivity\cite{sc03}. Moreover effective theories
can be derived for these extreme conditions, using macroscopic
degrees of freedom, by matching them to QCD at a scale close to
the chiral scale $\Lambda_\chi \sim 4\pi f_\pi \sim 1$
GeV\cite{HY03}.

We have followed a different path to dense matter studies by using
as our starting point a model Lagrangian, in the spirit of Skyrme,
which describes hadronic matter and meson dynamics respecting the
symmetries of QCD. The parameters of the model are fixed by meson
dynamics at {\em zero} baryon number density. {\it \`A la\/}
Skyrme\cite{Sk62}, baryons arise from a soliton solution, the
skyrmion, with the topological winding number describing the
baryon number. In our scheme dense matter is approximated by a
system of skyrmions with a given baryon number density whose
ground state arises as a crystal
configuration\cite{SkyrmionCrystal,KS88}. Starting from this
ground state our approach provides insight on the intrinsic
in-medium dependence of meson dynamics. We have studied (i) the
in-medium properties of the mesons and (ii) the role of the other
degrees of freedom besides pions in the description of matter as
it becomes
denser\cite{PMRV02,LPMRV03,LPRV03a,LPRV03b,PRV03,KPMRV04}.

\section{Model Lagrangians}
The original Skyrme model Lagrangian\cite{Sk62} reads
\begin{equation}
{L}_{\pi} =
- \frac{f_\pi^2}{4} \mbox{Tr} (L_\mu L^\mu)
+ \frac{1}{32e^2} \mbox{Tr} [ L_\mu, L_\nu]^2
+ \frac{f_\pi^2 m_\pi^2}{4} \mbox{Tr} (U + U^\dagger - 2),
\label{L:p}
\end{equation}
where $L_\mu = U^\dagger \partial_\mu U$ and
$U=\exp(i\vec{\tau}\cdot\vec{\pi}) \in SU(2)$ is a nonlinear
realization of the pion fields, $f_\pi$  the decay constant and
$m_\pi$ the pion mass . The second term with $e$, the Skyrme
parameter, was introduced to stabilize the soliton solution.

The dilaton field $\chi$ was incorporated in the model to make it
consistent with the scale anomaly of QCD\cite{EL85}. The
Lagrangian (\ref{L:p}) then becomes
\begin{equation}
\begin{array}{rcl}
{L}_{\pi\chi}
&=& \displaystyle
- \frac{f_\pi^2}{4} \left(\frac{\chi}{f_\chi}\right)^2
            \mbox{Tr} (L_\mu L^\mu)
+ \frac{1}{32e^2} \mbox{Tr} [ L_\mu, L_\nu]^2  \\
&& \displaystyle
+ \frac{f_\pi^2 m_\pi^2}{4} \left(\frac{\chi}{f_\chi}\right)^3
            \mbox{Tr} (U + U^\dagger - 2) \\
&&
+ \frac12 \partial_\mu \chi \partial^\mu \chi - V(\chi),
\end{array}
\label{L:pc}
\end{equation}
Note the different powers of $(\chi/f_\chi)$ in front of each
term. The last line is  the Lagrangian for the free dilaton field,
where $V(\chi)=(m_\chi^2 f_\chi^2/4)((\chi/f_\chi)^4
(\mbox{ln}(\chi/f_\chi)-\frac{1}{4})-\frac{1}{4})$, $m_\chi$ is
the dilaton mass and $f_\chi$ its decay constant.

The vector mesons, $\rho$ and $\omega$, can be incorporated into
the Lagrangian as dynamical gauge bosons of a hidden local gauge
symmetry which requires the doubling of the degrees of freedom as
$U=\xi^\dagger_L \xi_R$. This Lagrangian reads\cite{HLS}
\begin{equation}
\begin{array}{rcl}
{L}_{\pi\chi\rho\omega}
&=& \displaystyle
- \frac{f_\pi^2}{4} \left(\frac{\chi}{f_\chi}\right)^2
  \mbox{Tr}(L_\mu L^\mu)
+ \frac{f_\pi^2 m_\pi^2}{4} \left(\frac{\chi}{f_\chi}\right)^3
    \mbox{Tr}(U+U^\dagger-2)
\\
&& \displaystyle
-\frac{f_\pi^2}{4} a \left(\frac{\chi}{f_\chi}\right)^2
 \mbox{Tr}[ \xi_L \partial_\mu \xi_L^\dagger
          + \xi_R \partial_\mu \xi_R^\dagger
          + i(g/2)( \vec{\tau}\cdot\vec{\rho}_\mu + \omega_\mu)]^2
\\
&&
+ \frac{3}{2} g \omega_\mu B^\mu
\\
&&
-\textstyle \frac{1}{4} \displaystyle
\vec{\rho}_{\mu\nu} \cdot \vec{\rho}^{\mu\nu}
-\textstyle \frac{1}{4}  \omega_{\mu\nu} \omega^{\mu\nu}
+\textstyle\frac{1}{2} \partial_\mu \chi \partial^\mu \chi
-V(\chi),
\end{array}
\label{L:pcrw}
\end{equation}
with $\vec{\rho}_{\mu\nu} = \partial_\mu \vec{\rho}_\nu
 - \partial_\nu \vec{\rho}_\mu + g \vec{\rho}_\mu \times \vec{\rho}_\nu$,
$\omega_{\mu\nu}=\partial_\mu \omega_\nu -
\partial_\nu\omega_\mu$, and where $B_\mu$ is the {\em topological}
baryon number current. The quartic Skyrme term of (\ref{L:p}) is
not present, because its stabilizing role is played here by the
vector mesons.


\section{In-Medium Pion Dynamics}
In order to explain our basic strategy, we start with the Skyrme
model (\ref{L:p}). Table 1 serves to generalize the discussion to
other models. The vacuum solution of (\ref{L:p}) in the $B=0$
sector is $U=1$. Fluctuations on top of this vacuum describe pion
dynamics. The Lagrangian supports solitons with nontrivial
topological structures. The $B=1$ soliton solution with lowest
energy is a ``hedgehog" , $U^{B=1} = \exp(i \vec{\tau}\cdot\hat{r}
F(r))$ with the boundary condition for the profile function
$F(r)$, $F(0)=\pi$ and $F(\infty)=0$. The two-skyrmion system has
lowest energy when one on the skyrmions is rotated relatively to
the other in isospin space, by an angle $\pi$, about an axis
perpendicular to the line joining their centers. Therefore,
skyrmion matter has the lowest energy in an FCC (face centered
cubic) single skyrmion crystal structure, where the skyrmions at
the nearest site are relatively oriented in such a low energy
configuration and $U(\vec{r})$ has the symmetry structure
described in Table 1.

\begin{table}[b]
\caption{Summary of the properties of the  $B=0$ vacuum solution,the
hedgehog Ansatz for the $B=1$ skyrmion, the symmetries of the FCC
single skyrmion crystal for pions, dilaton and vector mesons.}
{\footnotesize
\begin{tabular}{ccc}
\hline
 & pion \& rho & dilaton \& omega \\[1ex]
\hline
vacuum
 & $U=1$, $\rho^a_\mu=0$
 & $\chi=f_\chi$, $\omega_\mu=0$  \\[1ex]
\hline
hedgehog
 & $\begin{array}{l}
      U=\exp(i\vec{\tau}\cdot\hat{r} F(r)) \\
      \rho^a_i = \varepsilon_{aip} \hat{r}_p G(r)/gr
   \end{array}$
 & $\begin{array}{c}
      \chi=\chi(r)\\
      \omega_0=\omega(r)
   \end{array}$  \\[1ex]
\hline
$\begin{array}{c}
  \mbox{boundary} \\  \mbox{condition}
\end{array}$
 & $\begin{array}{cc}
      F(0)=\pi & F(\infty)=0 \\
      G(0)=-2 & G(\infty)=0
    \end{array}$
 & $\begin{array}{cc}
      \omega^\prime(0)=0 & \omega(\infty)=0 \\
      \chi^\prime(0)=0 & \chi(\infty)=0
    \end{array}$ \\[1ex]
\hline
$\begin{array}{c}
   \mbox{reflection}^{(1)} \\ \mbox{($yz$-plane)}
 \end{array}$
 & $(\phi^{\pi,\rho}_0, -\phi^{\pi,\rho}_1,
      \phi^{\pi,\rho}_2, \phi^{\pi,\rho}_3)^*$
 & $\chi$, $\omega$ \\[1ex]
$\begin{array}{c}
   \mbox{3-fold axis} \\ \mbox{rotation}^{(2)}
 \end{array}$
 & $(\phi^{\pi,\rho}_0, \phi^{\pi,\rho}_2,
      \phi^{\pi,\rho}_3, \phi^{\pi,\rho}_1)$
 & $\chi$, $\omega$ \\[1ex]
$\begin{array}{c}
   \mbox{4-fold axis} \\ \mbox{rotation}^{(3)}
 \end{array}$
 & $(\phi^{\pi,\rho}_0, \phi^{\pi,\rho}_1,
      \phi^{\pi,\rho}_3, -\phi^{\pi,\rho}_2)$
 & $\chi$, $\omega$ \\ [1ex]
$\begin{array}{c}
   \mbox{FCC} \\ \mbox{translation}^{(4)}
 \end{array}$
 & $(\phi^{\pi,\rho}_0, -\phi^{\pi,\rho}_1,
      -\phi^{\pi,\rho}_2, \phi^{\pi,\rho}_3)$
 & $\chi$, $\omega$ \\ [1ex]
\hline
$\begin{array}{c}
   \mbox{Ansatz for} \\ \mbox{fluctuations}
 \end{array}$
 & $\begin{array}{c}
       U=\sqrt{U_\pi} U_M \sqrt{U_\pi} \\
       \rho_\mu^a = \rho^{a,M}_\mu + \tilde{\rho}_\mu^a
    \end{array}$
 & $\begin{array}{c}
       \chi=\chi^{}_M + \tilde{\chi} \\
       \omega_\mu = \omega^{}_{\mu,M} + \tilde{\omega}_\mu
    \end{array}$ \\ [1ex]
\hline
\multicolumn{3}{l}
{$^*{ }U=\phi^\pi_0 + i\vec{\tau}\cdot\vec{\phi}^\pi$,
$\rho^a_i = \varepsilon_{abc} \phi^\rho_b \partial_i
            \phi^\rho_b/(1+\phi^\rho_0)$} \\
\multicolumn{3}{l}
{ $(x,y,z) \rightarrow$ $(1)$ $(-x,y,z)$,
 $(2)$ $(y,z,x)$, $(3)$ $(x,z,-y)$, $(4)$ $(x+L,y+L,z)$ }  \\
\end{tabular}
}
\end{table}

For a given baryon number density, the lowest energy configuration
can be found numerically by varying the field values at points of
a discrete mesh\cite{SkyrmionCrystal} or by adjusting the
coefficients of the Fourier series expansion for the fields with
specific symmetries\cite{KS88}.

\begin{figure}[t]
\centerline{\epsfxsize=15cm \epsfbox{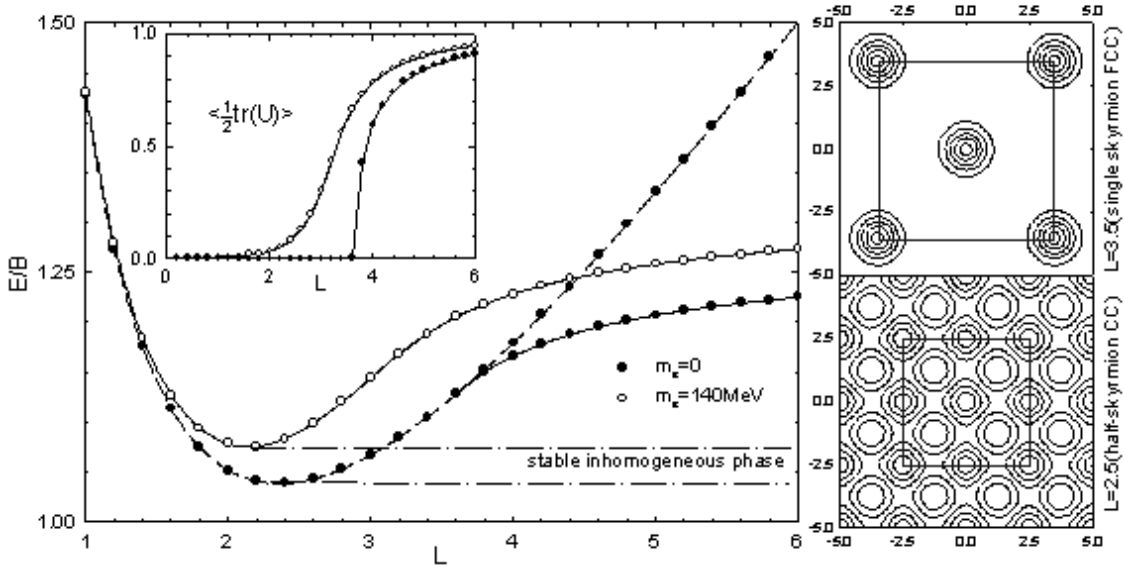}} 
\caption{This figure shows the energy per skyrmion and the average
value of $\sigma=\langle \frac12 \mbox{Tr}(U)\rangle$ as a
function of the FCC crystal parameter $L$ (in units of
$(ef_\pi)^{-1}$). The figures in the small boxes correspond to
samples of the baryon number distribution on the $xy$-plane
($z=0$). The corresponding FCC boxes are drawn by a square.}
\end{figure}

We show in Figure 1 the energy per baryon $E/B$ of skyrmion matter
and $\sigma$, the average value of $\frac12\mbox{Tr}U=\phi^\pi_0$
over space, as a function of the FCC single skyrmion crystal size
parameter $L$. Dimensionless units are used for $E/B$ (in units of
$6\pi^2 f_\pi/e$) and $L$(in units of $(ef_\pi)^{-1}$). For
massless pions, for a value of $L\sim 3.8$ the slope of $E/B$
changes indicating that the system undergoes a first order phase
transition. If we look into the actual baryon number distributions
(see figures in the small boxes), we see that the system changes
from the FCC single skyrmion crystal to a CC half-skyrmion
crystal. In the former phase, a well-localized single skyrmion is
located at each FCC lattice site where $\phi^\pi_0=-1$. In the
latter phase, one half of the baryon number carried by the single
skyrmion is concentrated at each FCC site while the other is
concentrated on the links where $\phi^\pi_0=+1$. Both
``half-skyrmions" centered at the points where $\phi^\pi_0=\pm 1$
can hardly be distinguished and therefore they form a CC crystal.
This phase transition can be seen more apparently in the quantity
$\langle \frac12 \mbox{Tr}U \rangle$ (see the inset figure). In
the literature\cite{SkyrmionCrystal}, the vanishing of $\sigma$ is
often interpreted as the restoration of the chiral symmetry.

If we turn on the pion mass (results shown in the figure by open
circles), we see no sudden change in the slope of $E/B$ but at
sufficiently high baryon number density skyrmion matter behaves as
in an approximate half-skyrmion phase.

The phase to the left of the minimum, referred in our work as
``homogeneous", is described by a crystal configuration. The phase
to the right of  the minimum, which we called ``inhomogeneous",
because the pressure $P\equiv\partial E/\partial V$ is negative
and therefore skyrmion matter is unstable against condensation of
the skyrmions into dense lumps leaving large volumes of space
empty.

We incorporate the pion fluctuations through the Ansatz
\begin{equation}
U(\vec{r},t) = \sqrt{U_\pi} U_M(\vec{r}) \sqrt{U_\pi},
\end{equation}
where $U_M(\vec{r})$ describes the background baryonic matter and
$U_\pi=\exp(i \tau_a \tilde{\pi}_a/f_\pi)$ describes the
fluctuating pions. Substitution of this Ansatz into (\ref{L:p})
leads to
\begin{equation}
\begin{array}{rcl}
{L}_\pi^{B \neq 0}
&=&
\frac12 G_{ab}(\vec{r}) \partial_\mu \tilde{\pi}^a \partial^\mu \tilde{\pi}_b
+ \frac14 m_\pi^2 \mbox{Tr}(U_M) \tilde{\pi}_a \tilde{\pi}_a \\
&&
- \partial_\mu \tilde{\pi}_a A^{\mu}_a(\vec{r})
- \varepsilon_{abc} \tilde{\pi}_a \partial_\mu \tilde{\pi}_b V^\mu_c(\vec{r})
+ \cdots ,
\end{array}
\label{LM:p}
\end{equation}
where we have expanded up to the second order in the fluctuating
fields. Eq.(\ref{LM:p}) describes the dynamics of the  pion  in
the dense medium. $G_{ab}(\vec{r})$, $V_\mu^a(\vec{r})$ and
$A_\mu^a(\vec{r})$ are the interaction potentials appearing due to
background matter.

As a first approximation let's average the potentials over space.
Due to the reflection symmetries of the background matter,
$\langle V_\mu^a \rangle = \langle A_\mu^a \rangle =0$, $\langle
G_{ab} \rangle = \langle 1-\phi^\pi_a \phi^\pi_b \rangle \equiv
Z_\pi^2 \delta_{ab}$ and $\langle \phi^\pi_0 \rangle \equiv
\sigma$. Thus, the Lagrangian becomes
\begin{equation}
{L}_\pi^{B \neq 0}
= \textstyle
\frac12  Z_\pi^2 \partial_\mu \tilde{\pi}^a \partial^\mu \tilde{\pi}_a
+ \frac12 m_\pi^{2} \sigma \tilde{\pi}_a \tilde{\pi}_a.
\label{LM:p:avrg}\end{equation}
The factor $Z_\pi$ in front of the kinetic term can be absorbed into the
renormalization of the pion fields as
\begin{equation}
\tilde{\pi}_a \rightarrow  \tilde{\pi}^\prime = Z_\pi \tilde{\pi}_a.
\end{equation}
In terms of this newly defined fields $\tilde{\pi}^\prime_a$, the
Lagrangian can be rewritten as
\begin{equation}
{L}_\pi^{B \neq 0}
= \textstyle
  \frac12 \partial_\mu \tilde{\pi}_a^\prime
  \partial^\mu \tilde{\pi}^\prime_a
+ \frac12 m_\pi^{2} (\sigma/Z_\pi^2)
  \tilde{\pi}^\prime_a \tilde{\pi}^\prime_a.
\end{equation}
This implies that the pions effective mass in the medium is given
by
\begin{equation}
m_\pi^* = m_\pi \sqrt{\sigma}/Z_\pi.
\label{mpi:star}\end{equation} Furthermore, we may reinterpret the
wave function renormalization factor $Z_\pi$ as the ratio between
the {\em in-medium} decay constant and the free space one,
\begin{equation}
f_\pi^*/ f_\pi = Z_\pi.
\label{fpi:star}\end{equation}

\begin{figure}[t]
\centerline{\epsfxsize=15cm \epsfbox{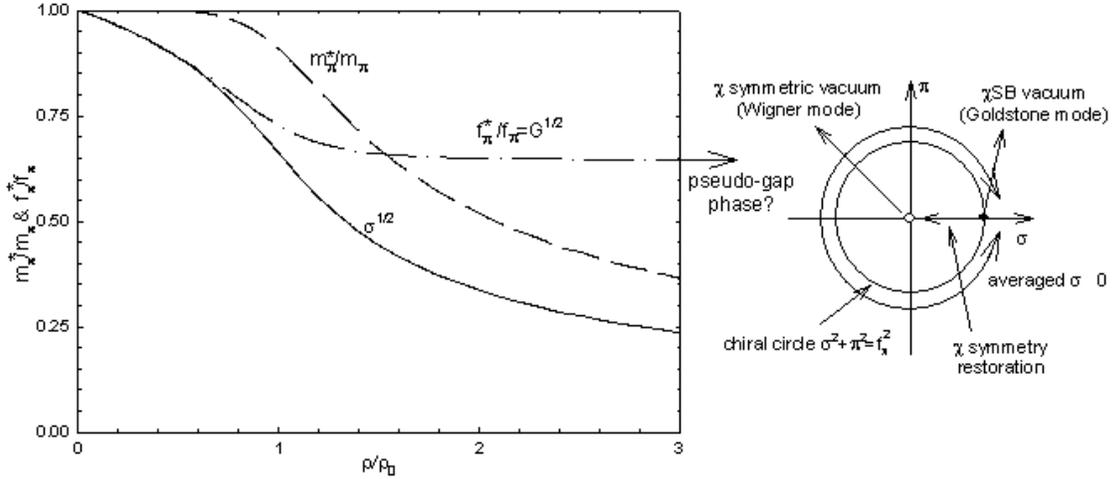}} 
\caption{We show the effective pion mass and decay constant in the
dense medium as a function of the baryon number density. The
dependence on the baryon number density approximately scales with
the Skyrme parameter $e$ as $\sim e^{-3}$. The right figure
illustrates a pseudo-gap scenario.}
\end{figure}

The results on these in-medium quantities (\ref{mpi:star}) and
(\ref{fpi:star}) are presented in Figure 2 as a function of the
baryon number density. The scales are strongly dependent on the
the parameters $f_\pi$ and $e$, e.g. the density units in $\rho$
could change considerably, and therefore the shown numbers should
not be taken as definitive. However, the qualitative behavior will
remain unchanged. In particular, the dependence on the baryon
number density showing approximate scaling with the Skyrme
parameter $e$ as $\sim e^{-3}$ is a solid statement.

In Figure 2, we show that at low densities $m_\pi^*/m_\pi \sim 1$,
while at higher density the ratio decreases down to zero. The
density dependence does not appear linear in $\rho$, since the
classical background leads automatically to higher powers in the
density dependence. At higher densities, powers greater than one
in $\rho$ come to play important roles. As the density increases
the decay constant $f^*_\pi$ decreases, which can be interpreted
as a partial restoration of the chiral symmetry in the medium.
However, the ratio $f_\pi^*/f_\pi$ only decreases up to $\sim
0.65$ remaining constant thereafter. This non vanishing of the
pion decay constant, despite the vanishing of $\sigma$, indicates
that we may be describing a phase which is not in the standard
Wigner-Weyl symmetry. It may be a pseudo-gap phase where the gap
is non-zero though chiral symmetry is restored, resembling what
might be happening in the normal phase of high $T_c$
superconductivity\cite{pseudogap}. The pseudo-gap phase is
schematically illustrated in Figure 2.

The pseudo-gap phase may be an artifact of the model with only
pions. If we have only pion degrees of freedom which are realized
non-linearly through the phase of $U$, they must live on the
chiral circle and the pseudo-gap phenomenon might appear.

However if we introduce a dilaton field $\chi$, as in
(\ref{L:pc}), we see that it may shrink the chiral circle to a
point and the conventional Wigner-Weyl symmetry appears
\cite{LPRV03a}. To illustrate this we take the $\chi$ field
constant, $\chi/f_\chi = X$. In baryon free space the vacuum of
the dilaton field is the minimum point of $V(\chi)$ and is $X=1$.
However, for the dense baryonic matter, the ground state should be
the minimum of $E/B$, the energy per baryon expressed as
\begin{equation}
E/B(X) = X^2 (E_2/B) + (E_4/B) + X^3 (E_m/B) + (2L^3)V(X).
\end{equation}
Here, $E_2$, $E_4$ and $E_m$ are the contributions from the
skyrmion field configurations of the given baryon number density
through the kinetic term, Skyrme term and mass term of the Skyrme
Lagrangian, respectively. Couplings to  dilaton matter terms
contributes with other effective potentials which will modify the
value of $X$ at which $E/B$ has a minimum further.

\begin{figure}[t]
\centerline{\epsfxsize=15cm \epsfbox{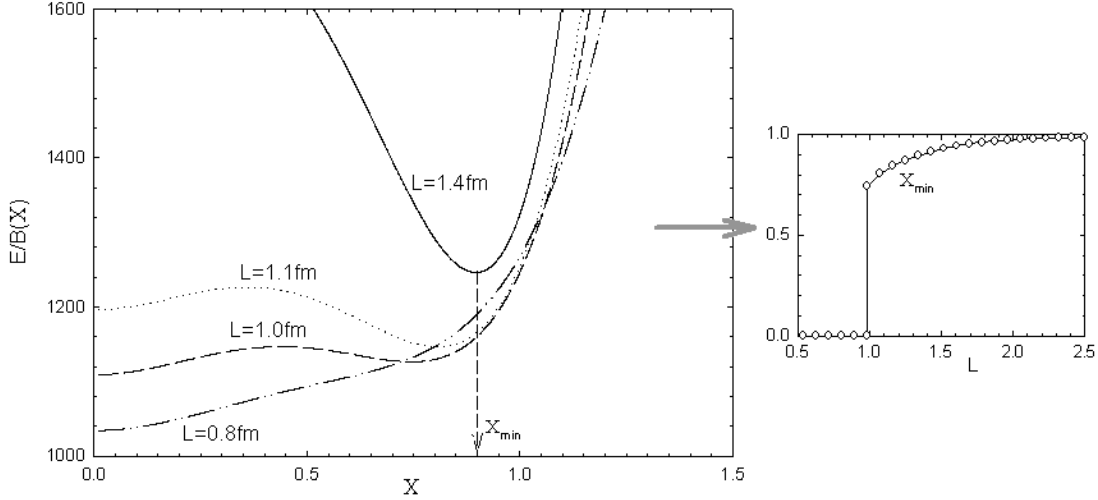}} 
\caption{The effective potential $E/B(X)$ as a function of the
constant scalar field $X$ for various $L$. }
\end{figure}

Shown in Figure 3 is $E/B$ as function of $X$ for various FCC
length parameters, where we have substituted for $E_2/B$, $E_4/B$
and $E_m/B$ the values that led to Figure 1. At low density (large
$L$), the minimum of the {\em effective} potential $E/B(X)$ is
shifted slightly away from $X=1$. As the density increases,
$E/B(X)$ starts developing another minimum at $X=0$ which was an
unstable extremum in free space. At $L\sim 1$ fm, the newly
developed minimum can compete with the one near $X\sim 1$. At
higher density, the minimum gets shifted to $X=0$. The figure in
the small box presents the resulting values of $X_{min}$ as a
function of $L$. There we see an explicit manifestation of a first
order phase transition. This mechanism corresponds to
reformulating BR-scaling\cite{BR91} in a more accurate way.

One can perform a more rigorous treatment by allowing the space
dependence in $\chi_M(\vec{r})$ (with the symmetries summarized in
Table 1), which leads basically to the same physics except for
small quantitative difference. The most essential new ingredient
is that the static dilaton field for dense matter vanishes
identically {\em all over the space} in the half-skyrmion phase.

By incorporating the fluctuations of pions and dilaton, as
summarized in Table 1, we obtain as before the Lagrangian for the
dynamics of these particles

\begin{equation}
L(U, \chi)={L}_{{\rm M}, \pi} + {L}_{{\rm M}, \chi}
          +{L}_{{\rm M}, \pi\chi},
\label{L:M:pc}
\end{equation}
where we have put the subscript ``M" to denote that it describes
the {\em in-medium} dynamics. Explicitly, each term can be
expressed as
$$ \begin{array}{rcl}
{L}_{{\rm M}, \pi} &=&
  \frac{1}{2} G_{ab}(\vec{r})
    \partial_\mu \tilde{\pi}_a \partial^\mu \tilde{\pi}_b
- \frac{1}{2} S(\vec{r}) \tilde{\pi}_a^2 \\
&& + \epsilon_{abc} \partial_\mu \tilde{\pi}_a \tilde{\pi}_b
    V^i_c(\vec{r})
\end{array}
\eqno(\mbox{\ref{L:M:pc}a})$$
$$
{L}_{{\rm M},\chi} =
  \frac{1}{2} \partial_\mu \tilde{\chi} \partial^\mu \tilde{\chi}
 -\frac{1}{2} M(\vec{r}) \tilde{\chi}^2,
\eqno(\mbox{\ref{L:M:pc}b})$$
$$
{L}_{{\rm M}, \chi\pi} = P_{a}^i(\vec{r}) \tilde{\chi} \partial^i
\tilde{\pi}_a + Q_a(\vec{r})  \tilde{\chi}\tilde{\pi}^a.
\eqno(\mbox{\ref{L:M:pc}c})$$ Here, $G_{ab}(\vec{r})$,
$S(\vec{r})$, $V_c^i(\vec{r})$, $M(\vec{r})$, $P_a^i(\vec{r})$ and
$Q_a(\vec{r})$ are the effective potentials provided to the
fluctuating fields by the background fields $U_M(\vec{r})$ and
$\chi_M^{}(\vec{r})$ (For the details, see Ref.\cite{LPRV03a}).

\begin{figure}[t]
\centerline{\epsfxsize=15cm \epsfbox{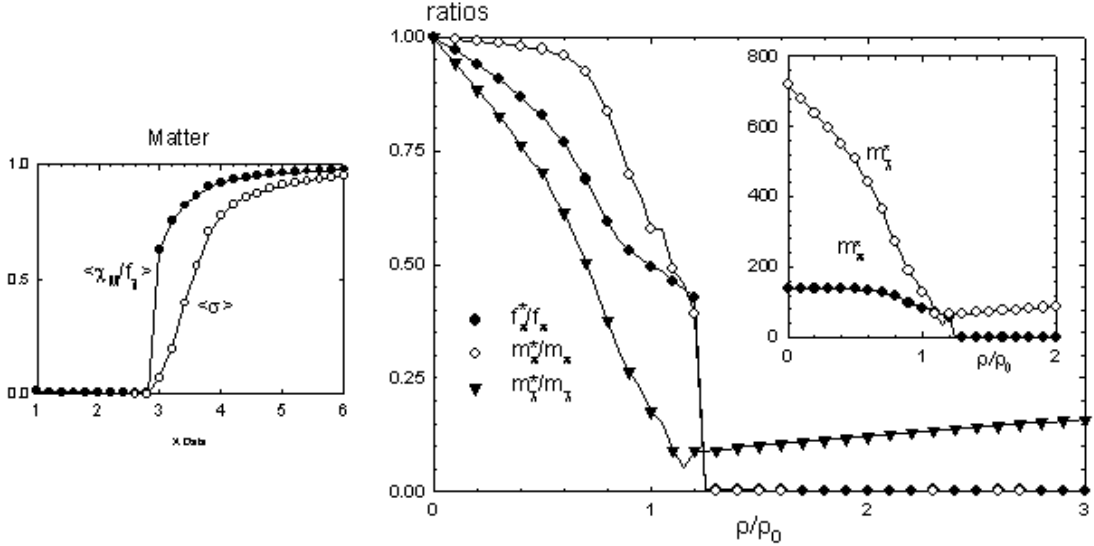}} 
\caption{In medium properties of the pion and dilaton fluctuations
as a function of the baryon number density. In the inset figure,
the pion and dilaton masses are presented in the actual energy
scale. The numerical results on the average values of $\chi_M^{}$
and $\sigma$ of the background matter are presented in the small
figure on the left hand side.}
\end{figure}

By applying the same approximation, i.e taking the average values
for the effective potentials over the space, we can estimate the
medium effects on the properties of the fluctuating fields. As for
the pion, we are led to a Lagrangian similar to (\ref{LM:p:avrg}),
where $Z_\pi$ and $\sigma$ take the average values with the
additional factors $(\chi^{}_M/f_\chi)^2$ and
$(\chi^{}_M/f_\chi)^3$, respectively. Since $\chi^{}_M$ vanishes
in the chiral symmetry restored phase, the ratio $f^*_\pi/f_\pi$
now vanishes. However, even in the  chirally restored phase, with
$\langle \sigma \rangle = \langle \chi_M^{} \rangle =0$, $\langle
M \rangle$ does not vanish due to the background $U_M(\vec{r})$
couplings to $\tilde{\chi}$. In Figure 4 we show the ratios of the
in-medium parameters relative to their free-space values.

The phenomenon discussed above is closely related to ``Brown-Rho"
scaling~\cite{BR91}. In the description of Ref.\cite{BR91}, the
density dependence comes solely from the change in the mean field
$\chi^*$ where the corresponding change in the skyrmion structure
has been ignored. Our present result corrects this fact and gives
a precise meaning to the scaling relation of Ref.\cite{BR91}.

We can treat the background interactions more systematically
following the {\em perturbative} scheme we have developed. We
decompose the Lagrangian into an unperturbed part, ${L}_0$, and an
interaction part, ${L}_I$. This leads to a Hamiltonian split also
in two parts
\begin{equation}
H=H_0 + H_I.
\end{equation}
The free propagators are defined by ${H}_0$ and the interaction
potentials appearing due to ${H}_I$ are summarized in Figure 5. In
Figure 5, $G^{ab}(\vec{\ell})$, for example, is the Fourier
transform of the local potential $G^{ab}(\vec{r})$:
\begin{equation}
G^{ab}(\vec{\ell}) = \frac{1}{V_{\mbox{\scriptsize box}} }
\int_{\mbox{\scriptsize box}} d^3 r e^{i\vec{\ell}\cdot\vec{r}}
G^{ab}(\vec{r}),
\end{equation}
where the integration is over a unit box of the crystal and
$V_{\mbox{\scriptsize box}}$ is its volume. Due to the periodic
structure of the crystal only discrete values of the momentum are
allowed.

\begin{figure}
\begin{center}
{\footnotesize \setlength{\unitlength}{0.7mm}
\begin{picture}(150,65)
\linethickness{0.8pt}
\put(29,62){\makebox(0,0)[b]{$\pi_a$}}
\put(28,60){\circle*{1}} \multiput(28,60)(5,0){7}{\line(1,0){3}}
\put(62,60){\circle*{1}}
\put(61,62){\makebox(0,0)[b]{$\pi_b$}}
\put(45,57){\makebox(0,0)[c]{$(p_0,\vec{p})$}}
\put(45,48){\makebox(0,0)[c]{$\displaystyle\frac{\delta_{ab}}
{p_0^2-\vec{p}^2}$}} \put(88,62){\makebox(0,0)[b]{$\chi$}}
\put(88,60){\circle*{1}} \put(88,60){\line(1,0){34}}
\put(122,60){\circle*{1}} \put(123,62){\makebox(0,0)[b]{$\chi$}}
\put(105,57){\makebox(0,0)[c]{$(p_0,\vec{p})$}}
\put(105,48){\makebox(0,0)[c]{$\displaystyle\frac{1}{p_0^2-\vec{p}^2
- m_\chi^2}$}}
\put(30,25){\makebox(0,0)[c]{${H}^{\pi\pi}_I(\vec{\ell})$}}
\put(20,26){\makebox(0,0)[b]{$\pi_a$}}
\put(23,24){\makebox(0,0)[tr]{$(p_0,\vec{p})$}}
\put(40,26){\makebox(0,0)[b]{$\pi_b$}}
\put(38,24){\makebox(0,0)[tl]{$(q_0,\vec{q})$}}
\put(23,25){\circle*{1}} \put(30,25){\circle{14}}
\put(37,25){\circle*{1}}
\put(30,13){\makebox(0,0)[c]{$-(p_0^2-\vec{p}\cdot\vec{q})
(G^{ab}(\vec{\ell})-\delta^{ab})$}} \put(30,7){\makebox(0,0)[c]
{$+i \epsilon_{abc}\vec{p}\cdot \vec{V}^c (\vec{\ell}) $}}
\put(75,25){\makebox(0,0)[c]{${H}^{\pi\chi}_I(\vec{\ell})$}}
\put(67,26){\makebox(0,0)[b]{$\chi$}}
\put(68,24){\makebox(0,0)[tr]{$(p_0,\vec{p})$}}
\put(85,26){\makebox(0,0)[b]{$\pi_a$}}
\put(83,24){\makebox(0,0)[tl]{$(q_0,\vec{q})$}}
\put(68,25){\circle*{1}} \put(75,25){\circle{14}}
\put(82,25){\circle*{1}}
\put(75,10){\makebox(0,0)[c]{$\displaystyle i
\vec{q}\cdot\vec{P}^a(\vec{\ell})$}}
\put(120,25){\makebox(0,0)[c]{${H}^{\chi\chi}_I(\vec{\ell})$}}
\put(112,26){\makebox(0,0)[b]{$\chi$}}
\put(113,24){\makebox(0,0)[tr]{$(p_0,\vec{p})$}}
\put(129,26){\makebox(0,0)[b]{$\chi$}}
\put(128,24){\makebox(0,0)[tl]{$(q_0,\vec{q})$}}
\put(113,25){\circle*{1}} \put(120,25){\circle{14}}
\put(127,25){\circle*{1}}
\put(120,10){\makebox(0,0)[c]{$M(\vec{\ell})-m_\chi^2$}}

\end{picture}
}
\end{center}
\caption{Free propagators and interactions for the pion and the
scalar fields in the presence of background skyrmion matter. The
energy-momentum conservation delta functions are not shown. }
\end{figure}

\begin{figure}
\begin{center}
{\footnotesize \setlength{\unitlength}{0.7mm}
\begin{picture}(150,60)(0,-5)
\linethickness{0.8pt}
\put(30,45){\makebox(0,0)[c]{$\Sigma^{(1)}$}}
\put(20,46){\makebox(0,0)[b]{$\pi_a$}}
\put(23,44){\makebox(0,0)[tr]{$(p_0,\vec{p})$}}
\put(40,46){\makebox(0,0)[b]{$\pi_a$}}
\put(38,44){\makebox(0,0)[tl]{$(p_0,\vec{p})$}}
\put(23,45){\circle*{1}} \put(30,45){\circle{14}}
\put(37,45){\circle*{1}}
\put(52,45){\makebox(0,0)[c]{=}}
\put(75,45){\makebox(0,0)[c]{${H}^{\pi\pi}_I(\vec{0})$}}
\put(65,46){\makebox(0,0)[b]{$\pi_a$}}
\put(68,44){\makebox(0,0)[tr]{$(p_0,\vec{p})$}}
\put(85,46){\makebox(0,0)[b]{$\pi_a$}}
\put(83,44){\makebox(0,0)[tl]{$(p_0,\vec{p})$}}
\put(68,45){\circle*{1}} \put(75,45){\circle{14}}
\put(82,45){\circle*{1}}
\put(30,25){\makebox(0,0)[c]{$\Sigma^{(2)}$}}
\put(20,26){\makebox(0,0)[b]{$\pi_a$}}
\put(23,24){\makebox(0,0)[tr]{$(p_0,\vec{p})$}}
\put(40,26){\makebox(0,0)[b]{$\pi_a$}}
\put(38,24){\makebox(0,0)[tl]{$(p_0,\vec{p})$}}
\put(23,25){\circle*{1}} \put(30,25){\circle{14}}
\put(37,25){\circle*{1}}
\put(52,25){\makebox(0,0)[c]{=}}
\put(75,25){\makebox(0,0)[c]{${H}^{\pi\pi}_I(\vec{\ell})$}}
\put(65,26){\makebox(0,0)[b]{$\pi_a$}}
\put(68,24){\makebox(0,0)[tr]{$(p_0,\vec{p})$}}
\put(85,26){\makebox(0,0)[b]{$\pi_b$}}
\put(83,24){\makebox(0,0)[tl]{$(p_0,\vec{p}+\vec{\ell})$}}
\put(68,25){\circle*{1}} \put(75,25){\circle{14}}
\put(82,25){\circle*{1}} \multiput(82,25)(4,0){5}{\line(1,0){2.5}}
\put(108,25){\makebox(0,0)[c]{${H}^{\pi\pi}_I(-\vec{\ell})$}}
\put(98,26){\makebox(0,0)[b]{$\pi_b$}}
\put(118,26){\makebox(0,0)[b]{$\pi_a$}}
\put(116,24){\makebox(0,0)[tl]{$(p_0,\vec{p})$}}
\put(101,25){\circle*{1}} \put(108,25){\circle{14}}
\put(115,25){\circle*{1}}
\put(52,5){\makebox(0,0)[c]{+}}
\put(75,5){\makebox(0,0)[c]{${H}^{\pi\chi}_I(\vec{\ell})$}}
\put(65,6){\makebox(0,0)[b]{$\pi_a$}}
\put(68,4){\makebox(0,0)[tr]{$(p_0,\vec{p})$}}
\put(85,6){\makebox(0,0)[b]{$\chi$}}
\put(83,4){\makebox(0,0)[tl]{$(p_0,\vec{p}+\vec{\ell})$}}
\put(68,5){\circle*{1}} \put(75,5){\circle{14}}
\put(82,5){\circle*{1}} \put(82,5){\line(1,0){20}}
\put(108,5){\makebox(0,0)[c]{${H}^{\chi\pi}_I(-\vec{\ell})$}}
\put(98,6){\makebox(0,0)[b]{$\chi$}}
\put(118,6){\makebox(0,0)[b]{$\pi_a$}}
\put(116,4){\makebox(0,0)[tl]{$(p_0,\vec{p})$}}
\put(101,5){\circle*{1}} \put(108,5){\circle{14}}
\put(115,5){\circle*{1}}
\end{picture}
}
\end{center}
\caption{Diagrams used to evaluate the self-energy of the
$\pi_a$ propagation up to second order in the interaction.
Here, $b$ runs over $1,2,3$ and the intermediate states run over
all $\vec{\ell} \neq 0$.}
\end{figure}

We show in Figure 6 the diagrams used to evaluate the self-energy.
Only the diagrams for $\Sigma_{\pi_a \pi_b}$ appear. The symmetry
structure of skyrmion matter allows a non vanishing self-energy
only for $a=b$. To first order, $\Sigma^{(1)}$ is nothing but
${H}_I(\vec{\ell}=\vec{0})$. Since ${H}_{\pi\chi}(\vec{0})=0$, no
mixing between the fluctuating pions and the fluctuating scalar
occurs. Thus, the pion propagator for $\tilde{\pi}_a$ can be
expressed as
\begin{equation}
\frac{1}{p_0^2- \vec{p}^2 - \Sigma^{(1)}(p_0,\vec{p})} =
\frac{1}{G^{aa}(\vec{0}) (p_0^2 - \vec{p}^2 ) }.
\label{propagator}
\end{equation}
where we have used that the self
energy to this order is $ \Sigma^{(1)}_{\pi_a\pi_a}(p_0,\vec{p})=
-p^2 (G_{aa}(\vec{0})-1). $ The superscript ``(1)" means that the
quantities are evaluated to first order.

Since Lorentz symmetry is broken by the medium, the general form
of the {\em in-medium} propagator can be written as
\begin{equation}
\frac{1}{Z_t^{-1} p_0^2 - Z_s^{-1}\vec{p}^2},
\label{prop_matter}
\end{equation}
 with $Z_{t,s}^{-1}=(f_{t,s}/f_\pi)^2$ and the ``pion velocity" in
medium is given as $v_\pi^2 = Z_t/Z_s$. Comparing with this,  we
obtain $f_t=f_s=f_\pi\sqrt{G_{aa}(\vec{0})}$. Since
$G_{aa}(\vec{0})$ is nothing but the average of $G^{aa}(\vec{r})$
over the space, our calculation thus far reproduces the naive
approximations discussed above. To the same order, the self-energy
of the scalar field is $\Sigma^{(1)}_{\chi\chi} =
M(\vec{0})-m_\chi^2$. Since it is constant, this self-energy
modifies just the scalar mass from the free value $m_\chi$ to
$\sqrt{M(\vec{0})}$.

Now, the second order diagrams shown in Figure 6 can be calculated
similarly. Again, in spite of the $\pi-\chi$ coupling term in the
interaction Lagrangian (\ref{L:M:pc}), $\Sigma^{(2)}_{\chi\pi_a}$
vanishes so that the pion propagator and the scalar propagator can
be simply written as
\begin{equation}
\frac{1}{p_0^2 - \vec{p}^2 - \Sigma^{(1+2)}_{\pi_b\pi_a}}, \hskip
3em \frac{1}{p_0^2 - \vec{p}^2 - m_\chi^2 -
\Sigma^{(1+2)}_{\chi\chi}},
\end{equation}
respectively.

The $p_0$ and $\vec{p}$ dependence of $\Sigma^{(2)}$ is not so
simple as that of $\Sigma^{(1)}$. By assuming that energy and
momentum are
 small, we may expand the self-energy $\Sigma_{\pi_a \pi_b}$ in
powers of $p_0$ and $\vec{p}$. Then, we can express the propagator
in the form of (\ref{prop_matter}), but $Z_s$ gets some
corrections terms from the second order diagram. The corrections
are negative definite so that the pion velocity becomes  $v_\pi <
1$.

\begin{figure}[t]
\centerline{\epsfxsize=11cm \epsfbox{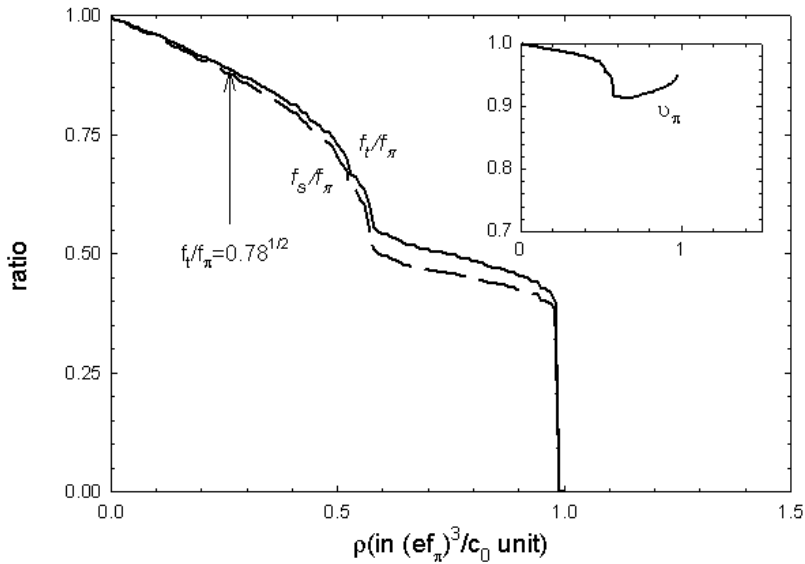}} 
\caption{$f_t$ and $f_s$ as a function of skyrmion matter density
in arbitrary units for the chiral limit, $m_\pi=0$ and
$m_\chi=720\ {\rm MeV}$. The in medium pion velocity appears as
function of matter density in the small box. The arrow indicates
the point at which the observed ratio in pionic nuclei (at
$\rho\approx 0.6\rho_0$) is located. } \end{figure}

The results of our calculation are shown in Figure 7. We show the
decay constants $f_t$ and $f_s$ in units of $f_\pi$ as a function
of the density measured in a dimensional units. The inset figure is
their ratio, i.e. the in-medium pion velocity $v_\pi$. To second
order the contribution to $f_s/f_\pi$ turns out to be small, and
thus the pion velocity stays $v_\pi \sim 1$. The lowest value is
about $\sim 0.9$. Furthermore, for the pions at higher matter
densities, the internal propagator provides an extra suppression
because $L$ scales as $\rho^{1/3}$. Once we pass the density at
which the pion velocity has its minimum, the pion velocity
increases with density and approaches 1. When the background
matter is in the half-skyrmion phase, $\chi_0(\vec{r})$ vanishes
identically and so do all the local potentials. Thus, both $f_t$
and $f_s$ vanish. Moreover, their difference is proportional to
the square of the potentials and vanishes faster. This fact
explains the asymptotic behavior $v_\pi \rightarrow 1$ in Figure
7. This result is very similar to that of Ref.~\cite{HKRS03} found
in heat bath, where the pion velocity approaches 1 while both the
spatial and temporal pion decay constants vanish at $T=T_c$.


\section{Vector mesons}

At higher density and/or temperature we need more degrees of
freedom such as the $\rho$ and $\omega$ vector mesons. The same
procedure can be repeated with the vector mesons using the
Lagrangian (\ref{L:pcrw}). The lowest energy FCC skyrmion crystal
configuration can be found by requiring the symmetries given in
Table 1 to the pion, dilaton, rho and omega fields. The numerical
results are presented in Figure 8.

\begin{figure}[t]
\centerline{\epsfxsize=15cm \epsfbox{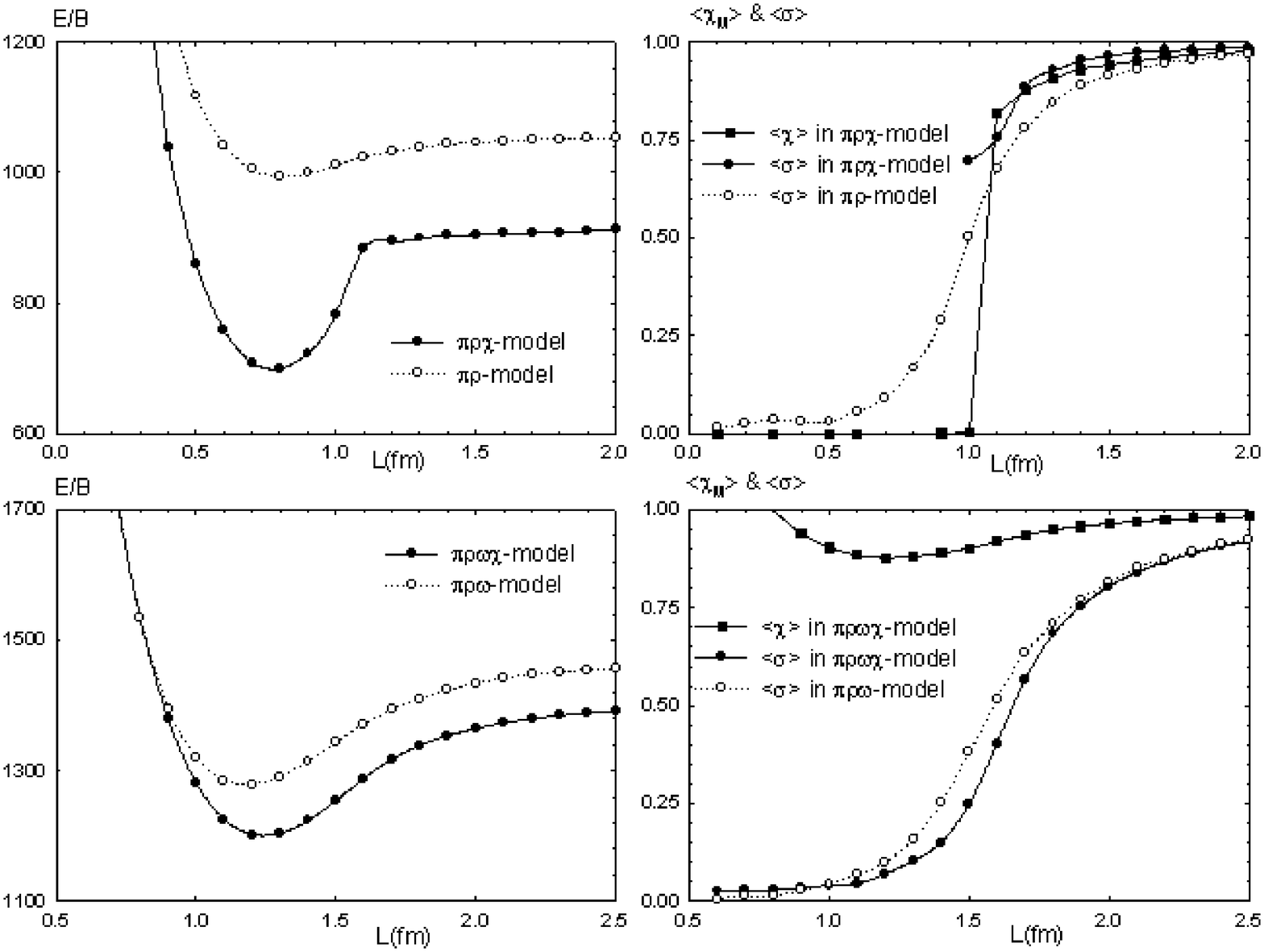}} 
\caption{$E/B$ and the average values of the fields over the space
as a function of $L$. \label{pcrw}}
\end{figure}

In the $\pi \rho \chi$ model, as the density of the system
increases ($L$ decreases), $E/B$ changes slightly. Its value is
close to the energy of a single skyrmion up to densities larger
than $\rho_0\; (L \sim 1.43)$. because the size of the skyrmion is
very small and the skyrmions in the lattice only interact at very
high densities when their tails overlap.

In the absence of the $\omega$ the dilaton field plays as before
an important role. Skyrmion matter undergoes an abrupt phase
transition at the density at which the expectation value of the
dilaton field vanishes $\langle \chi \rangle=0$.

In the  $\pi \rho \omega \chi$ model, the situation changes
dramatically, because the $\omega$ provides not only a strong
repulsion between the skyrmions but also an intermediate range
attraction. In both the $\pi\rho\omega$ and the
$\pi\rho\omega\chi$ models, at high density, the interaction
reduces $E/B$ to 85\% of the $B=1$ skyrmion mass. This value
should be compared with 94\% in the $\pi\rho$ model. In the
$\pi\rho\chi$-model, $E/B$ goes down to 74\% of the $B=1$ skyrmion
mass, but in this case it is due to the dramatic behavior of the
dilaton field.

In the $\pi\rho\omega\chi$ model the role of the dilaton field is
suppressed. It provides only a small attraction at intermediate
densities. Moreover, the phase transition towards its vanishing
expectation value, $\langle \chi\rangle=0$, does not take place.
Instead, its value grows at high density!

The reason for this can be found in the role played by omega in
(\ref{L:pcrw}). In the static configuration, omega produces a
potential, whose source is the baryon number density, which
mediates the self-interaction energy of the baryon number
distribution. Thus, unless it is screened properly by the omega
mass, the periodic source filling infinite space will lead to an
infinite self-energy. To reduce the energy of the system, the
effective $\omega$ mass must grow at high density, for which
$\chi$ must grow too. Note the factor $(\chi/f_\chi)^2$ in the
omega mass term in Lagrangian (\ref{L:pcrw}).

\section{Summary}
We have developed a unified approach to dense matter in the Skyrme
philosophy, where systems of baryons and mesons can be described
by a single Lagrangian. In our approach dense baryonic matter is
approximated by skyrmion matter in the lowest energy configuration
for a given baryon number density. By incorporating in it
fluctuating mesons we can get some insights on meson dynamics in a
dense medium. Our approach enables us to study this dynamics
beyond the first order in the baryon number density. One can
continue to work in this direction by incorporating more degrees
of freedom, by improving the way of treating matter beyond the
crystal solution, and so on.

However, before closing the presentation, we must clearly lay down
the scope of our work. We do not claim that the results obtained
at present describe reality. The most fundamental problem we phase
is that our ``ground state" for matter is a crystal not a Fermi
liquid. Our aim has been to assume a state for matter, given by a
classical solution of a theory considered to be valid at large
$N_c$, and have studied the implications for its excitations. Our
work should be taken as representing the first step towards a more
realistic treatment of a dense matter theory.

\section*{Acknowledgments}
H.-J. Lee, B.-Y. Park and V. Vento are grateful for the
hospitality extended to them by KIAS. This work was partially
supported by grants MCyT-FIS2004-05616-C02-01 and
GV-GRUPOS03/094~(VV) and KOSEF Grant R01-1999-000-00017-0~(BYP \&
DPM).


\end{document}